# Observation of magnon-polarons in a uniaxial antiferromagnetic insulator

Junxue Li[1], Haakon T. Simensen[2], Derek Reitz[3], Qiyang Sun[4], Wei Yuan[1], Chen Li[4], Yaroslav Tserkovnyak[3], Arne Brataas[2], and Jing Shi[1]

1. Department of Physics and Astronomy, University of California, Riverside, California 92521, USA

2. Center for Quantum Spintronics, Department of Physics, Norwegian University of Science and Technology, NO-7491 Trondheim, Norway

3. Department of Physics and Astronomy, University of California, Los Angeles, California 90095, USA

4. Materials Science and Engineering/Department of Mechanical Engineering, University of California, Riverside, CA 92521, USA

Magnon-polarons, a type of hybridized excitations between magnons and phonons, were first reported in yttrium iron garnet as anomalies in the spin Seebeck effect responses. Here we report an observation of antiferromagnetic (AFM) magnon-polarons in a uniaxial AFM insulator $Cr_2O_3$. Despite the relatively higher energy of magnon than that of the acoustic phonons, near the spin-flop transition of ~ 6 T, the left-handed magnon spectrum shifts downward to hybridize with the acoustic phonons to form AFM magnon-polarons, which can also be probed by the spin Seebeck effect. The spin Seebeck signal is founded to be enhanced due to the magnon-polarons at low temperatures.

Magnon-polarons are hybridized excitations emerged from magnons and phonons due to the magnetoelastic coupling (MEC) in magnetic materials [1-8]. Analogous to polaritons formed from the photon and optical phonon hybridization in semiconductors [9], magnon-polarons modify the dispersion of both magnons and phonons and thus affect the thermodynamic and transport properties of the materials. Theoretically proposed over six decades ago [10-14], magnon-polarons have not been confirmed until recently in the spin Seebeck effect (SSE) and inelastic neutron scattering (INS) studies on ferrimagnetic insulators including yttrium iron garnet (YIG) [15-19] and $Lu_2BiFe_4GaO_{12}$ [20]. Unlike INS which requires large single crystal samples, SSE exploits pure spin currents carried by magnons; therefore, it is capable of probing small magnon-polaron anomalies in thin films.

Antiferromagnetic (AFM) materials have recently been experimentally demonstrated as a coherent or incoherent pure spin current source excited resonantly or thermally [21-25]. In addition, spin currents in AFMs can play a similar role to those in ferromagnets (FMs) in carrying angular momentum and delivering spin-orbit torques to manipulate the AFM Néel vector [26]. Understanding how magnon-phonon coupling and magnon-polarons affect the generation and transport of spin currents in AFMs becomes increasingly important to antiferromagnetic spintronics. In general, AFM magnon dispersion lies above that of acoustic phonons [27]. One exception is the non-collinear AFMs in which the low-lying magnon excitation modes can hybridize with the acoustic phonon modes. The magnon-acoustic phonon coupling in such non-collinear AFMs as $TbMnO_3$ and $Y(Lu)MnO_3$ was investigated by INS [28, 29], but the effect of the magnon-acoustic phonon hybridization on spin current generation and transport in these materials have not been explored. In collinear AFM materials, due to the higher magnon energy, the hybridization only takes place with optical phonons [30]. For example, such hybridization was predicted to be at 11.3 THz (~ 47.0 meV) and 17.3 THz (~ 72.0 meV) in NiO, but the frequency range is too high to be accessible by thermodynamic or transport means.

In this Letter, we report an observation of the AFM magnon-polarons arising from the hybridization of magnons and acoustic phonons in a uniaxial AFM insulator $Cr_2O_3$. The combination of the relatively low magnon energy (~ 0.696 meV at the Brillouin zone center) compared to other uniaxial AFMs and that we can further reduce the energy of one magnon branch to zero by applying a 6 T magnetic field, ensures that we can induce hybridization of magnons with both longitudinal acoustic (LA) and transverse acoustic (TA) phonons.

Figure 1(a) shows the magnon energy $E$ vs. the easy-axis magnetic field $H$ for the uniform spin precession mode ($k = 0$) of the uniaxial AFM insulator $Cr_2O_3$. Below the spin-flop (SF) field of 6 T, there are two distinct branches corresponding to the two eigen-modes (modes 1 and 2) of AFM magnons [31-33], *viz.* right-hand (RH) and left-hand (LH) circular spin precessions with opposite chiralities, as depicted in Fig. 1(b). At $H = 0$, these two modes are degenerate with the energy of $E_0 = 0.696$ meV and carry equal but opposite angular momenta, $\pm\hbar$. When $H$ is applied along the *c*-axis of $Cr_2O_3$, the degeneracy between modes 1 and 2 is lifted, and their dispersions are given in Ref. 34. This is markedly different from that of FMs. First, there is only one magnon mode in FMs, i.e., the right-handed one. Second, while $E_0$ is determined by the exchange and anisotropy energies for AFMs, it is only determined by the anisotropy energy in FMs [35], which makes the magnon dispersion of AFMs generally much higher than that of FMs. As $H$ reaches the SF transition field $H_{SF}$, spins in both sublattices rotate abruptly to be nearly perpendicular to $H$ with a small inclination. Modes 1 and 2 are replaced by two new modes [34, 36], i.e., modes 3 and 4 (Fig. 1(b)). In mode 3, the Néel vector $\boldsymbol{l} = \boldsymbol{m_1} - \boldsymbol{m_2}$ is linearly polarized but the net magnetic moment $\boldsymbol{m} = \boldsymbol{m_1} + \boldsymbol{m_2}$ is elliptically polarized and precesses around $H$ with the RH chirality, where $\boldsymbol{m_1}$ and $\boldsymbol{m_2}$ are magnetizations of two spin sublattices. Mode 4 is characterized by both linearly polarized $\boldsymbol{l}$ and $\boldsymbol{m}$, and thus does not carry angular momentum.

Based on the exchange parameters and group velocities of acoustic phonons reported in the previous literatures [37-40], we plot the dispersions of magnon and acoustic phonons of $Cr_2O_3$ in Fig. 1(c) (see details in Supplemental Material, Note 1). Now we use these magnon and phonon dispersions to demonstrate the formation of AFM magnon-polarons. In the absence of a magnetic field, the AFM magnon dispersion lies well above those of acoustic phonons; therefore, no hybridization of magnons and acoustic phonons occurs in the entire Brillouin zone. With a finite $H$ applied along the *c*-axis, the dispersions of modes 1 and 2 shift in the opposite directions in energy. The dispersion for mode 2 continuously shifts down with increasing $H$ and intersects those of LA and TA phonons before reaching the horizontal axis at the SF transition (Fig. 1(c)). In the presence of MEC, magnon-phonon hybridization takes place at the crossing points, leading to AFM magnon-polarons (as illustrated in Fig. 1(d)) at certain wavevectors $k$'s and energies $E$'s.

Just as in YIG [15-19], the AFM magnon-polarons are expected to produce anomalies in SSE at specific magnetic fields corresponding to the touching points where the two dispersions are tangential to each other and the overlap regions are maximized in the *E-k* space.

Figure 2(a) shows the SSE measurement geometry for a $Cr_2O_3$/Ta heterostructure. We perform SSE measurements with an on-chip heater to generate a vertical temperature gradient. The open-circuit voltage is recorded as the SSE signal while **H** is swept along the *c*-axis of $Cr_2O_3$ (see details in Supplemental Material, Note 2). We normalize the SSE voltages by the heating power in order to fairly compare the effects for different measurement conditions (Supplemental Material, Note 3). Figure 2(c) displays the SSE data measured in $Cr_2O_3$/Ta at 2.2 K. The most salient feature in this field dependence is the abrupt jump at 6 T, the SF transition [23], which is consistent with our magnetic moment data (Note 4, Supplemental Material). The SSE sign change at $\mu_0 H_{SF}$ is correlated with the magnon mode switching from mode 2 to mode 3, which is accompanied with spin polarization switching [21, 34]. Below the SF transition, there is a large antisymmetric field-dependent SSE signal. After excluding the origin of the ordinary Nernst effect in the heavy metal layer (Supplemental Material, Note 3), we conclude that it is from the AFM magnon SSE mechanism [21, 34]. These features are in contrast with the absence of the SSE signal below the SF transition previously reported in $Cr_2O_3$ [23] and the absence of the SSE sign change across the SF transition in $MnF_2$ [24].

Another interesting feature in Fig. 2(c) is a pair of small but reproducible wiggles that stand out of the smooth SSE background right below $\mu_0 H_{SF}$ (Supplemental Material, Note 5). The field range of these wiggles (3.3-6.0 T) coincide with what is expected for the magnon-polaron anomalies, as illustrated in Fig. 1(c). These fine SSE structures are field-antisymmetric and exist over a range of temperatures as will be discussed below. To properly extract the SSE anomalies due to the magnon-polarons, we first fit the smooth part of the SSE signal below $H_{SF}$ using the theoretical model for SSE of uniaxial AFMs [34] (also see Supplemental Material, Note 6). The best fit is shown as the red curve in Fig. 2(c), which serves as the SSE background without anomalies. The wiggles represent enhanced SSE magnitude with respect to the background. After background subtraction, the two anomalies are separated out and plotted as $V_{mp}$ in the top panel of Fig. 2(e). Clearly, the anomalies are antisymmetric about the magnetic field. The anti-symmetry is expected for magnon-polarons because they are composed of magnons that should reverse the direction of the angular momentum or spin polarization as the magnetic field is reversed. To confirm the magnon-polaron origin of the anomalies, we replace Ta by Pt (as shown in Fig. 2(b)), both of which use the inverse spin Hall effect (ISHE) for SSE voltage detection but have the opposite signs in their spin Hall angles. Both the anomalies and SSE background signals are inverted, as shown in Figs. 2(d) and 2(e) (bottom panel). Therefore, we attribute the anomalies to the maximum hybridization between the magnon and both LA and TA phonons at $\mu_0 H_{m,LA}$ and $\mu_0 H_{m,TA}$ where the dispersions of magnon and phonons are tangential to each other (as illustrated in Fig. 2(f)). The positions of $\mu_0 H_{m,LA}$ and $\mu_0 H_{m,TA}$ are identified by the dips/peaks in $V_{\text{mp}}$ and indicated by the blue and red arrows in Fig. 2(e), respectively, and they agree quite well with the values expected from the magnon and phonon dispersions (Note 1 of Supplemental Material). The magnon-polaron is clearly absent above the SF transition in our data which can be explained by the vanished MEC strength (see details in Supplemental Material, Note 7).

To investigate the evolution of magnon-polaron signals as a function of temperature, we perform SSE measurements on $Cr_2O_3$/Ta at various temperatures from 2.2 K to 250 K. After properly subtracting the smooth SSE background for all curves (Supplemental Material, Note 6), we extract the magnon-polaron contributions and plot them in Fig. 3. The blue and red dashed lines in Fig. 3 mark the positions of the anomalies at $\pm\mu_0 H_{m,LA}$ and $\pm\mu_0 H_{m,TA}$ which are nearly independent of temperature, indicating the weak temperature dependence of magnon and phonon dispersions over this temperature range [37-40]. In addition, the magnon-polaron signal $V_{mp}$ depends linearly on the heating power (Note 8 in Supplemental Material). Fig. 3 shows a clear $V_{mp}$ minimum at $+\mu_0 H_{m,TA}$ at each temperature in spite of a rising background signal especially at high temperatures. The stronger positive background at high temperatures may have resulted from some high-order effects that are not included in the description of the AFM magnon model. Nevertheless, the local minimum persists at all temperatures before it vanishes. To track the magnitude of both anomalies as a function of temperature, we measure the depth of the dip in reference to the SSE value on the low-field side of the peak. For the LA feature, the low-field plateau serves as a good reference since no dispersion crossing exists and no SSE anomaly is expected. The TA anomaly, however, starts to emerge in the field range where the LA anomaly still remains finite which is caused by the two crossing points between the LA phonon and magnon dispersions (as discussed in Note 6 in Supplemental Material). Therefore, to evaluate the TA anomaly strength $V_{mp}$ at $+\mu_0 H_{m,TA}$, we choose the maximum signal between $+\mu_0 H_{m,LA}$ and $+\mu_0 H_{m,TA}$ as the reference point because it is where the overlap of the two anomalies is the smallest. Fig. 4(a) displays the temperature dependence of both the SSE background and magnon-polaron signals at $+\mu_0 H_{m,LA}$ and $+\mu_0 H_{m,TA}$. Here we point out that the negative $V_{mp}$ in $Cr_2O_3$/Ta represents an enhancement of the SSE signal, i.e., the greater total SSE signal magnitude than that of the smooth background. To better assess the temperature dependence of $V_{mp}$, we normalize it by the SSE background signal $V_{SSE}^{bg}$ at the same magnetic fields to eliminate the common thermal conductivity effect on both, as presented in Fig. 4(b). Clearly, $V_{mp}/V_{SSE}^{bg}$ signals at $+\mu_0 H_{m,LA}$ and $+\mu_0 H_{m,TA}$ decrease quickly and vanish below 6 K. On the negative field side, although $V_{mp}$ holds the opposite sign at $-\mu_0 H_{m,LA}$ and $-\mu_0 H_{m,TA}$, the same conclusion can be drawn as that for the positive field side (Supplemental Material, Note 9). We also observe qualitatively the same temperature dependence of the $V_{mp}$ signals in the Pt device (Supplemental Material, Note 10). We should point out that our conclusions are based on the SSE background analysis that was discussed earlier. Different SSE backgrounds adopted for subtraction may result in quantitatively different magnon-polaron signal magnitude.

In previous studies on YIG, the enhancement or suppression of the SSE signal caused by the magnon-polarons was found to depend on the relative strength of the magnetic and non-magnetic impurity scattering potentials [15,16,20], parameterized by $\eta = |v^{mag}/v^{ph}|^2$ [16], where $v^{mag}$ and $v^{ph}$ are isotropic impurity scattering potentials of magnons and acoustic phonons, respectively. When magnons are more strongly scattered than phonons ($\eta > 1$), the formation of the magnon-polarons leads to a longer magnon relaxation time and thus an enhancement in SSE magnitude. Here we assume that a similar mechanism works for AFM magnon-polarons. Since we observe enhanced $V_{mp}$ signals at both $\mu_0 H_{m,LA}$ and $\mu_0 H_{m,TA}$, it indicates that AFM magnons suffer stronger scattering than phonons do. In YIG, magnons were found to have a shorter mean-free-path than phonons [41, 42] at low temperatures, as a result of stronger magnon-impurity and magnon-magnon scatterings while the phonons already freeze out. A similar low-temperature scenario may also be true for $Cr_2O_3$. Actually, the signature of magnon-polarons can also be observed in the $Cr_2O_3$/Pt heterostructure with etched $Cr_2O_3$ surface, whereas the SSE signal magnitude below the SF transition is greatly suppressed. Both facts may be caused by stronger magnetic impurity scattering at the etched interface (Supplemental Material, Note 11).

In summary, we have observed SSE anomalies in a uniaxial AFM insulator $Cr_2O_3$ right below the SF transition and attributed the anomalies to the AFM magnon-polarons due to the hybridization of magnons and acoustic phonons. By tracking the temperature dependence of the SSE anomalies, we find that the magnon-polarons show similar behaviors between the LA and TA phonons. The enhanced SSE signal due to the presence of magnon-polarons indicates stronger magnon scattering at low temperatures. Our study demonstrates a unique capability of using SSE for investigating the interaction between magnons and phonons in AFM materials.

We acknowledge useful discussions with Igor Barsukov. Work at UC Riverside was supported as part of the Spins and Heat in Nanoscale Electronic Systems (SHINES), an Energy Frontier Research Center funded by the U.S. Department of Energy, Office of Science, Basic Energy Sciences under award no. SC0012670 (JXL and JS).

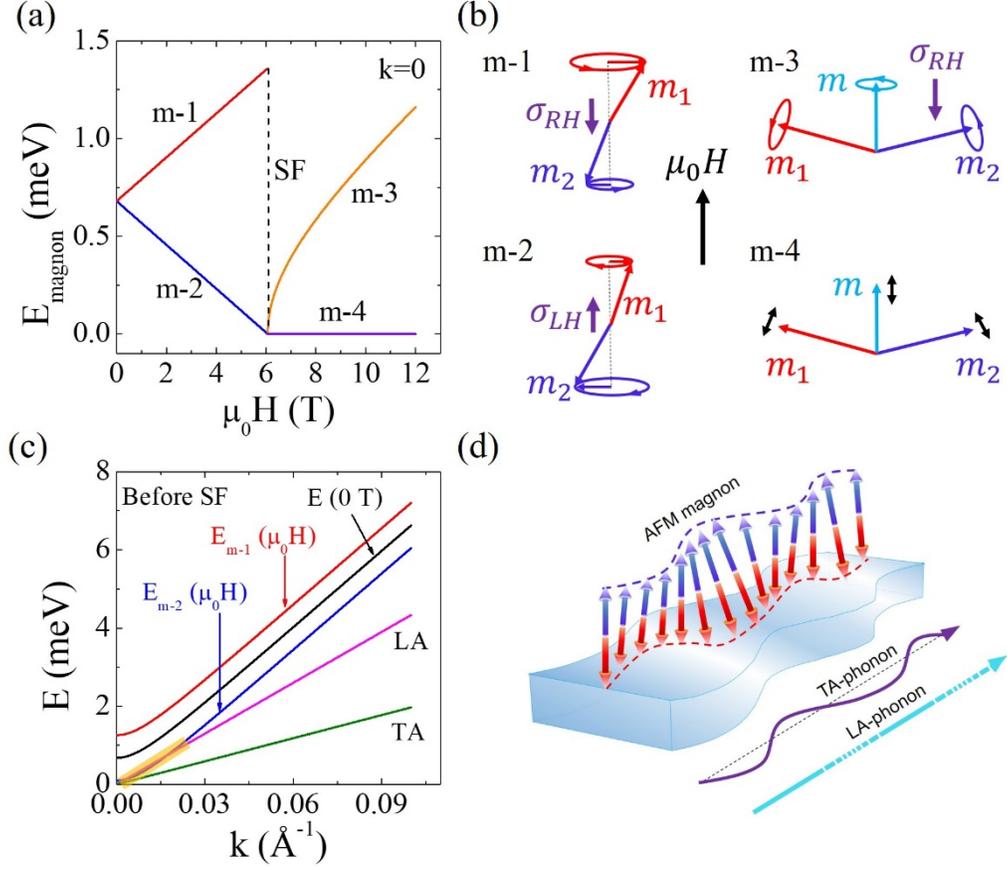

FIG. 1. Magnon-phonon hybridization in a uniaxial antiferromagnet. (a) Magnon energy of $Cr_2O_3$ at $k=0$ vs. magnetic field along the easy axis of $Cr_2O_3$. SF denotes the spin-flop transition. m-i (i=1,2,3,4) indicates AFM magnon mode i. (b) Four different AFM magnon modes under magnetic field $\mu_0 H$. $m_1$ and $m_2$ are the magnetizations of the two spin sublattices. modes 1 and 3 are the precessing modes with the right-hand (RH) chirality, whereas mode 2 is with the left-hand (LH) chirality, and mode 4 is linearly polarized. $\sigma_{RH}$ and $\sigma_{LH}$ are the spin polarizations associated with the RH and LH chiralities, respectively. (c) The magnon and acoustic phonon dispersions of $Cr_2O_3$ before SF. $k$ is perpendicular to the $(10\bar{1}0)$-plane. The black curve is the magnon dispersion at zero magnetic field, the red (blue) solid curve is the dispersion of magnon mode 1 (mode 2) at 5 T. Magnon mode 2 starts to hybridize with the acoustic phonons at high magnetic fields, as indicated by the thick yellow line segment. (d) Schematic real-space diagram of magnon-polaron in a uniaxial antiferromagnet.

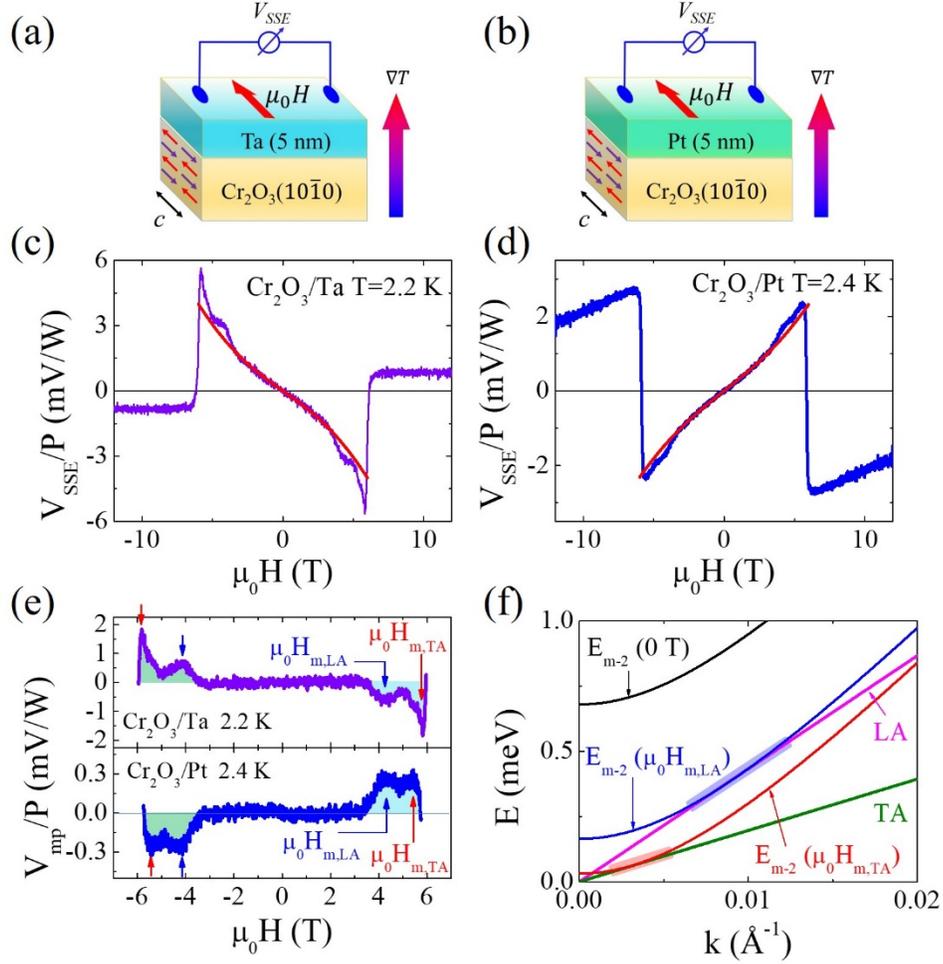

FIG. 2. Magnon-polaron SSE signals in $Cr_2O_3$/heavy metal heterostructures at low temperatures. (a) and (b), Device schematics for SSE measurements in $Cr_2O_3(10\bar{1}0)$/Ta(5 nm) and $Cr_2O_3(10\bar{1}0)$/Pt (5 nm), respectively. External field $\mu_0H$ is along the c-axis. (c) and (d), SSE signals in $Cr_2O_3$/Ta (at 2.2 K) and $Cr_2O_3$/Pt (at 2.4 K), respectively. The SSE signal is normalized to heating power $P$. The red curves in (c) and (d) are fitting results using eq. (S2a) in Supplemental Material, Note 6. (e) Magnon-polaron signals $V_{mp}/P$ in $Cr_2O_3$/Ta (at 2.2 K) and $Cr_2O_3$/Pt (at 2.4 K) obtained by subtracting the background from the original SSE signals in (c) and (d). The blue (red) arrows indicate the maximum hybridization between magnon and LA (TA) phonon. (f) Enlarged dispersions under magnetic fields of 0 T, $\mu_0H_{m,LA}$, $\mu_0H_{m,TA}$. The magnon dispersion under $\mu_0H_{m,LA}$ ($\mu_0H_{m,TA}$) becomes tangential to the dispersion of LA (TA) phonon.

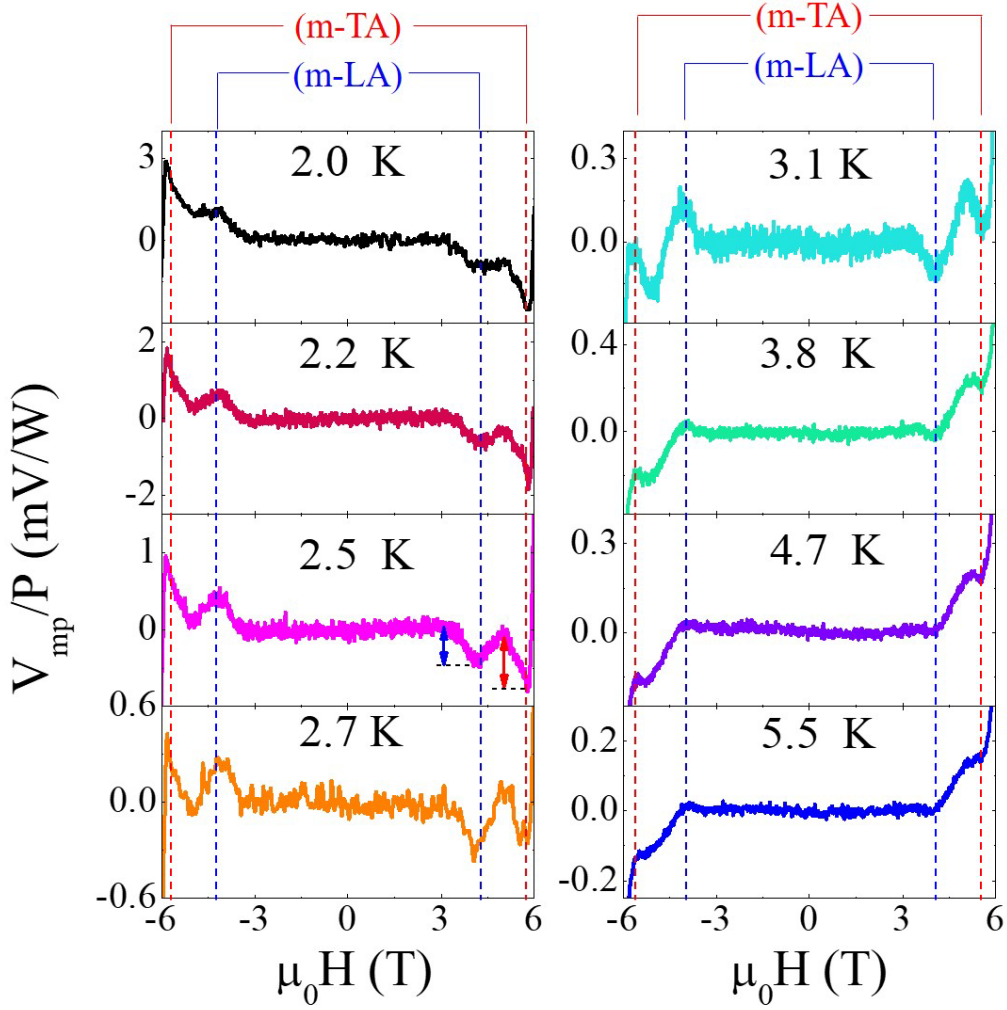

FIG. 3. Field dependence of magnon-polaron signals in $Cr_2O_3$/Ta heterostructure at different temperatures. $V_{mp}/P$ curves are obtained as discussed in Supplemental Material, Note 6. The blue and red dashed lines denote the magnetic fields $\pm\mu_0 H_{m,LA}$ and $\pm\mu_0 H_{m,TA}$ (as defined in Figs. 2(e) and 2(f)), respectively. The blue and red arrows in the 2.5 K panel indicate the magnitude of magnon-polaron signals for LA and TA phonon, respectively.

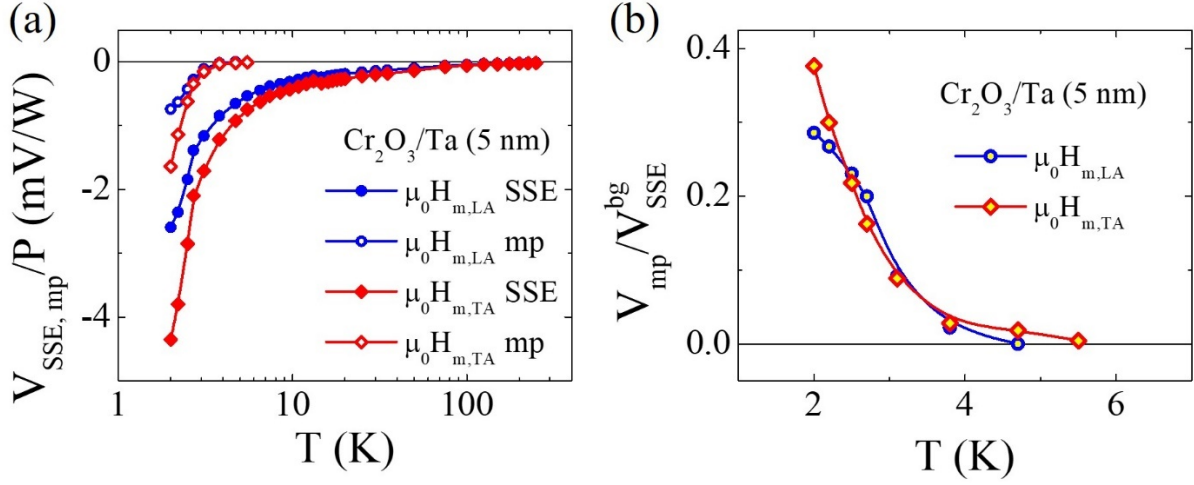

FIG. 4. Temperature dependence of magnon-polaron signals in $Cr_2O_3$/Ta heterostructure. (a) Magnon-polaron (mp) signal $V_{mp}/P$ and SSE background signal $V_{SSE}^{bg}/P$ at $+\mu_0 H_{m,LA}$ and $+\mu_0 H_{m,TA}$ vs. temperature. $V_{SSE}^{bg}/P$ is evaluated from the fitting results and contains no magnon-polaron contributions. (b) Temperature dependence of both LA-mp and TA-mp signals $V_{mp}$ normalized by the SSE background $V_{SSE}^{bg}$. The normalization eliminates the effect of the thermal conductivity on the temperature gradient which is common to both $V_{mp}/P$ and $V_{SSE}^{bg}/P$.